\documentstyle[aps,multicol,pra,epsfig]{revtex}
\begin{document}
\draft
\title{Dynamical Superfluid-Insulator
Transition\\ in a Chain of Weakly Coupled Bose-Einstein Condensates}
\author{A. Smerzi$^1$, A. Trombettoni$^{1,2}$, 
P.G. Kevrekidis$^3$ and A.R. Bishop$^1$}
\address{
$^1$ Theoretical Division and Center for Nonlinear Studies, Los Alamos
National Laboratory, Los Alamos, NM 87545, USA \\
$^2$ Istituto Nazionale di Fisica per la Materia and International School
for Advanced Studies, via Beirut 2/4, I-34014, Trieste, Italy \\
$^3$  Department of Mathematics and Statistics, University of 
Massachusetts, Amherst, MA 01003-4515, USA}
\date{\today}
\maketitle
\begin{abstract} 

We predict a dynammical classical superfluid-insulator transition (CSIT) in 
a Bose-Einstein condensate (BEC) trapped in an optical and a magnetic
potential. In the tight-binding limit, 
this system realizes
an array of weakly-coupled condensates driven by an external harmonic field.
For small displacements of the parabolic trap about the 
equilibrium position, 
the BEC center of mass oscillates 
with the relative phases of 
neighbouring condensates locked at the same (oscillating) value. 
For large displacements, the
BEC remains localized on the side of the harmonic trap. This is caused 
by a randomization of the relative phases,
while the coherence of each individual condensate in the array 
is preserved.
The CSIT
is attributed to a discrete modulational instability, occurring when 
the BEC center of mass velocity is larger 
than a critical value, 
proportional to the tunneling rate between adjacent sites.
\end{abstract}
\begin{multicols}{2}
The recent experimental investigations of the dynamical properties
of a Bose-Einstein condensate trapped in optical potentials 
\cite{anderson98,orzel01,morsch01,cataliotti01}, have led to
a rapidly growing interest in this topic 
\cite{chiofalo00,wu01,trombettoni01,konotop01,jaksch98,pedri01}.
The spatial and temporal coherence of matter waves emitted at 
different heights of the gravitational field 
has been proven in \cite{anderson98}, after loading a condensate
in a vertical optical trap. Number squeezed (non-classical) states 
have been realized in \cite{orzel01}. In \cite{morsch01} 
Bloch oscillations and interband transitions in an accelerating
lattice have been observed.
In \cite{cataliotti01}, the optical potential was superimposed on a
harmonic magnetic trap,  
realizing a chain of weakly coupled condensates 
(i.e., a Josephson junction array)
driven by an external parabolic field.
For small initial displacements, the condensate center of mass oscillated
symmetrically with the relative phases among adjacent
sites locked togheter.

Here we demonstrate that for large displacements, 
when the velocity of the center of mass reaches a critical value 
(proportional to the tunneling rate beteen adjacent sites),
the BEC abruptly stops
on the side of the harmonic trap (i.e., without reaching its center). 
We define an order parameter
for the system, and show that this dynamical transition from a
``superfluid'' to an ``insulator'' regime 
is associated with a randomization of the relative
phases among different wells.
As we will discuss below, 
this transition has a {\it classical} (mean-field or Gross-Pitaevskii) nature, 
and it is different
(but with some analogies) from the {\it quantum}
superfluid-insulator (Mott) 
transition caused by the number squeezing of the quantum states in each well 
\cite{jaksch98}.
It also differs  from the Landau dissipation mechanism, occurring 
in (quasi-)homogeneous systems when the
velocity of the condensate is larger than the sound speed \cite{wu01}.
Rather, the CSIT is driven by 
a modulational instability (MI) that causes 
an exponential growth of small perturbations of a 
carrier wave, as a result of the interplay between dispersion 
and nonlinearity.
The MI is a general feature of
discrete as well as continuum nonlinear wave equations.
Its demonstrations span a diverse set of disciplines ranging
from fluid dynamics \cite{benjamin67} (where it is usually referred
to as the Benjamin-Feir instability) 
and nonlinear optics \cite{ostrovskii69}
to plasma physics \cite{taniuti68}.
One of the early contexts in which its significance
was appreciated was the linear stability analysis 
of deep water waves. 
It was only much later recognized that the
conditions for MI would be significantly modified for discrete 
settings relevant to, for instance,
the local denaturation of DNA \cite{peyrard93} 
or coupled  arrays of optical waveguides 
\cite{morandotti99}. 
In the latter case,
the relevant model is the discrete nonlinear Scr\"odinger equation (DNLS), 
and its MI conditions were discussed in \cite{kivshar92}. 
In this letter we propose an experiment to observe a superfluid-insulator 
mean-field dynamical transition (as a consequence of the MI), 
with weakly coupled Bose-Einstein condensates 
driven by an external harmonic field.  

The BEC dynamics is governed by 
the Gross-Pitaevskii equation (GPE): 
\begin{equation}
\label{GPE}
i \hbar \frac{\partial \Phi}{\partial t}= - \frac{\hbar^2}{2 m} 
\nabla^2 \Phi + [V_{ext} + g_0 \mid \Phi \mid^2 -\mu] \Phi 
\end{equation}
where $g_0=\frac{4 \pi \hbar^2 a }{m}$,
$a$ is the $s$-wave scattering length, $m$ the atomic mass and
$\mu$ the chemical potential. 
The condensate wave function $\Phi(\vec{r},t)$ is normalized 
to the total number of condensate atoms $N_T$, and we consider a repulsive
interatomic interaction $a>0$. 
The external potential $V_{ext}$ 
is given by the sum of the harmonic confining potential 
$V_M=\frac{m}{2}[\omega_x x^2+
\omega_r^2(y^2+z^2)]$  
and the optical potential $V_L=V_0 \cos^2{(kx)}$. 
The valleys of the potential 
are separated by a ``lattice spacing'' of $\lambda/2$, with $\lambda=2\pi/k$. 
We consider a chemical potential $\mu<<V_0$ and the   
transverse degrees of freedom to be frozen by a tight magnetic
confinement, so as to justify the study of the system 
in an effective one-dimensional geometry.
 
In the tight-binding approximation $\Phi(\vec{r},t)=
\sqrt{N_T} \sum_n \psi_n(t) \phi_n(\vec{r})$, with the wavefunction
$\psi_j$ of the condensate in the $j$-th site of the array, weakly coupled
in the barrier region with 
the wavefunctions $\psi_{j \pm 1}$ of the condensates in the neighbour sites. 
It is then possible to map the GPE onto the DNLS 
\cite{trombettoni01}: 
\begin{equation}
\label{DNLS}
i  \hbar \frac{\partial \psi_n}{\partial t} = - K
(\psi_{n-1}+\psi_{n+1}) + (\epsilon_n+ U \mid \psi_n \mid ^2)\psi_n
\end{equation}
with $K \simeq - \int d\vec{r} \, \big[ \frac{\hbar^2}{2m} 
(\vec{\nabla} \phi_n \cdot \vec{\nabla} \phi_{n+1}) + \phi_n V_{ext} \phi_{n+1}
 \big] $
proportional to the microscopic tunneling rate between adjacent sites,
$U = g_0 N_T \int d\vec{r} \phi_n^4 $ and 
 $\epsilon_n=\int d\vec{r} [\frac{\hbar^2}{2m} 
 (\vec{\nabla} \phi_n)^2 + V_{ext} \phi_n^2]=\Omega n^2$, 
 with $\Omega=
\frac{1}{2} m \omega_x^2 \big(\frac{\lambda}{2} \big)^2$. 
The DNLS Hamiltonian is
\begin{equation}
\label{HAM}
{\cal{H}} =
{\sum_n} [ - K ( \psi_n \psi^\ast_{n+1} + \psi^\ast_n
\psi_{n+1} )
+ \epsilon_n \mid
\psi_n\mid^2 + {U \over 2} \mid\psi_n\mid^4  ]
\end{equation}
with $i \psi_n^\ast, \psi_n$ canonically conjugate variables.
Both $\cal H$ and the norm
$\sum_n \mid \psi_n \mid ^2 = 1$
are integrals of the motion.

Let us consider, first, the case $\epsilon_n=0$ (which corresponds to
neglecting the effect of the harmonic trap).
Stationary solutions of Eq. (\ref{DNLS}) 
are plane waves 
$\psi = \psi_0  \exp[i (k n - \nu t)]$, of
frequency
$\nu = - 2 K \cos(k) + U |\psi_0|^2$. 
The stability analysis of such states can be carried out by 
perturbing the
carrier wave with small amplitude phonons:
$\psi_n = \psi_0 e^{i (k n - \nu t)} + 
u e^{i (q n - \omega t)} + v^{\ast} e^{-i (q n + \omega t)}$.
The DNLS excitation spectrum (for $\epsilon_n=0$) is then given by:
$$
\omega_{\pm} = 2 K \sin(k) \sin(q) \pm 
$$
\begin{equation}
2
\sqrt{4 K^2 \cos^2(k) \sin^4({q \over 2}) + 
2 K U |\psi_0|^2 \cos(k) \sin^2({q \over 2})}.
\label{eq10}
\end{equation}
The carrier wave becomes modulationally unstable when  
the eigenfrequency $\omega$ in Eq.(\ref{eq10})
becomes imaginary:
\begin{equation}
 U |\psi_0|^2 > - 2 K \cos(k) \sin^2(q/2),
\label{eqn5}
\end{equation}
namely, when $\cos(k)<0$.
Therefore, if the interatomic interaction is repulsive ($U > 0$),
the system suffers an exponential growth of perturbations 
when  $\pi/2 < k < 3 \pi/4$.
This result will remain valid in the case
of non-homogeneous travelling wave-packets
driven by external fields, 
when their width is much larger than the wave length
associated with the collective motion. This conclusion can be further
understood in the light of the collective coordinate
equations of motion developed in \cite{trombettoni01}.
Generally speaking, the mapping of the GPE into the 
DNLS allows for the study of 
solitons and localized excitations  
as well as dynamical instabilities 
in the framework of the lattice theory 
\cite{scott99,hennig99,review01}. 
We remark, however, that the MI is a general feature of the GPE 
with a periodic 
external potential, and not necesserly in the tight binding limit.
In the perturbative limit, $\mu >> V_0$, and in absence of external driving
fields (i.e., with $V_M=0$), 
the MI has been studied in \cite{wu01,konotop01}
(of course, in this limit the MI condition doffers from Eq.(\ref{eqn5})). 

\begin{figure}
\centering
{\epsfig{file=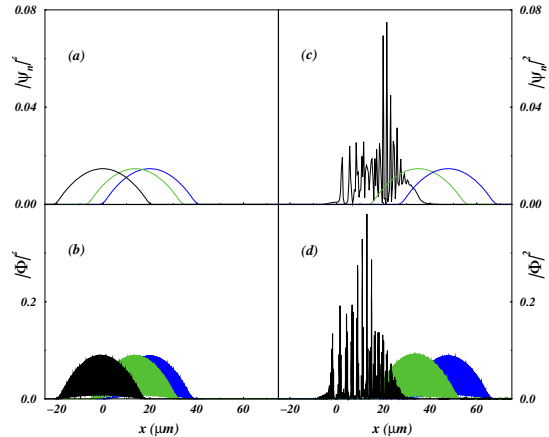, width=6.cm,angle=270, clip=}}
\caption{
The density calculated at different times $0$, $20$, $40$ $ms$ (from the
right of each figure to the left) as a function of the position, 
with initial displacements $\xi_(0)=50,120$ 
sites, which are, respectively, below and above 
the critical value $xi_{cr} \approx 84$ (\ref{cr-displ}). The GPE 
(b,d), and the DNLS (a,c) wavefunctions normalized to $1$
are compared.
(a,b) $\xi_(0)=50$ sites;
(c,d) $\xi_(0)=120$ sites.
The external parabolic potential, which drives the oscillations,
is centered at $x=0$.}
\label{fig3}
\end{figure}

The effect of the exponential growth 
of phonon modes of arbitrary momenta in the DNLS
leads to an effective dephasing among different sites of the lattice.
Indeed the phases of each condensate enter into a ``running regime'', 
with an angular velocity different from site to site and proportional to 
the local (on-site) effective chemical potential.
The complete delocalization in momentum space leads to 
a strong localization in real space, hence to the appearance
of localized structures of large amplitude (see Fig. \ref{fig3}).
This localization has also been attributed (in the absence of any external
potential $\epsilon_n = 0$) to the presence of
the so-called Peierls-Nabarro barrier \cite{peyrard84}, which pins such large
amplitude solutions \cite{dauxois93},
not allowing them to propagate.
The excess kinetic energy is partially stored
to high-frequency internal ``ac'' oscillations among 
adjacent wells (see also \cite{kivshar98,kevrekidis00b}), 
and partially converted to wakes of small amplitude
extended wave radiation \cite{peyrard84,kevrekidis00}.

The CSIT can be observed experimentally
by condensing, firstly, the 
atomic gas in both the magnetic and the optical trap, and, then,  
adiabatically displacing the magnetic field from its initial
position.
For small displacements, in line with the findings of \cite{cataliotti01},
the system coherently oscillates about the center
of the potential. 
If we rewrite $\psi_j=\sqrt{n_j} e^{i\phi_j}$, 
this implies that the phase difference 
between sites is given by
$\phi_{j+1}(t)-\phi_j(t) = \Delta \phi(t)$. 
The center of mass $\xi=\sum_j j n_j$ and the phase difference 
$\Delta \phi$ will then satisfy
\begin{eqnarray}
&&\hbar \frac{d}{dt}\xi(t) = 2K ~ \sin{\Delta \phi(t)}   \cr
&&\hbar \frac{d}{dt}\Delta \phi(t)= - 2 ~ \Omega ~ \xi(t).
\label{phase-current} 
\end{eqnarray}
Eqs. (\ref{phase-current}) have the usual form of the 
Josephson equations \cite{barone} and indicate that the overall 
array of bosonic 
Josephson junctions behaves as a {\em single} Josephson junction, whose 
critical current is $2K/\hbar$. 
The collective coherence was experimentally
demonstrated in \cite{cataliotti01} 
by the interference pattern obtained upon releasing
the condensates from the optical and magnetic traps.

\begin{figure}
\centerline
{\epsfig{file=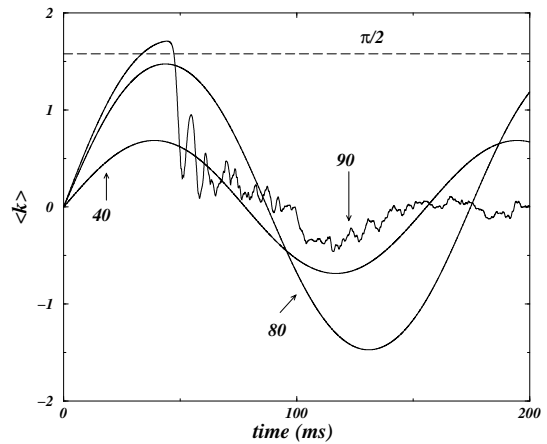, width=6.cm,angle=270}}
\caption{
The quasi-momentum $<k>$ vs. time for 
three different initial displacements: $40$, $80$ 
and $90$ sites.
When $<k>$ reaches $\pi/2$ 
(i.e., for an initial displacement greater 
than $\xi_{cr} \approx 84$ calculated with Eq.(\ref{cr-displ})
the system becomes modulationally unstable.} 
\label{fig1}
\end{figure}

\begin{figure}
\centering
{\epsfig{file=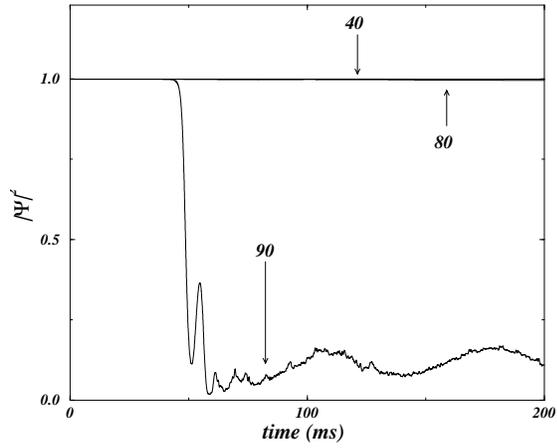, width=6.cm, angle=270, clip=}}
\caption{
The modulus square of the order parameter $\Psi$ defined
in Eq. (\ref{order-parameter}) as a function of time,
for three different
initial displacements ($40$, $80$ and $90$ sites) and
with the same parameters as in Fig. \ref{fig1}.
When the quasi-momentum
$<k>$ reaches $\pi/2$ (i.e.,
for an initial displacement greater than $\xi_{cr}$),
the order parameter drops to $\sim 0$; 
cf. Fig. \ref{fig1}.}
\label{fig2}
\end{figure}

To monitor the dynamical transition of interest, we define
$<k>=\sum_k k |\tilde{\psi}_k |^2= \Delta \phi$, 
with $\tilde{\psi}_k$ 
the Fourier transform of the condensate wave-function.
In the coherent, small amplitude oscillations regime,
the quasi-momentum $<k>$ exhibits
regular oscillations (see Fig. \ref{fig1}).
However, for 
$<k> \geq \pi/2$, the system becomes modulationally unstable
and localization ensues. The critical initial displacement 
$\xi_{cr}$ can therefore be obtained from Eqs. (\ref{phase-current})
with $\Delta \phi = \pi/2$. In lattice units: 
\begin{equation}
\label{cr-displ}
\xi_{cr}=\sqrt{\frac{2K}{\Omega}}.
\end{equation} 
In Fig. \ref{fig1} we plot $<k>$ vs. time for three initial displacements.
When $\xi(t=0)$ is smaller than $\xi_{cr}$, the average momentum $<k>$ 
oscillates in accordance with Eq. (\ref{phase-current}). Note that for
$\xi(t=80)$ the system approaches very close to the instability line.
When the initial displacement is larger than the critical value, 
$<k>$ abruptly drops as soon as it crosses the critical point. This is 
accompanied by the sudden arresting of the BEC center of mass (cf. 
Fig. \ref{fig3}) and by
a collective dephasing (cf. Fig. \ref{fig2}). 
The key experimental signature would be
the disappearance of the interference fringes after the expansion of
the BEC, while the
center of mass of the resulting cloud will be 
resting on the side of the trap's center.

\begin{figure}
\centering
{\epsfig{file=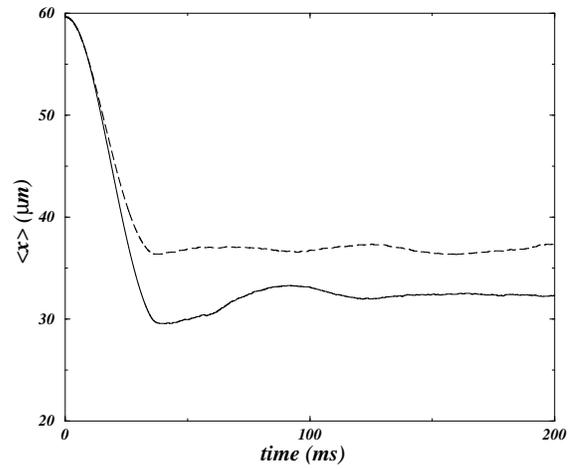, width=6.3cm,angle=270, clip=}}
\caption{The 
center of mass vs. time for a supercritical
initial displacement of $150$ sites. 
Solid line: Gross-Pitaevskii equation; dashed line: 
discrete nonlinear Schr\"odinger equation.
The (displaced) center of the trap is at $x = 0~ \mu$m.}
\label{fig4}
\end{figure}

The values of the parameters in DNLS are 
$V_0=5 E_R$ with 
$\lambda=795 \, nm$ and the recoil energy $E_R =\frac{h^2}{2m \lambda^2}$,
$N_T=50000$, $K=5.5 \cdot 10^{-2}E_R$ 
and $\Omega=1.5 \cdot 10^{-5}E_R$: 
the critical displacement from Eq. (\ref{cr-displ}) is 
$\xi_{cr} \approx 84$ sites, in good agreement with our numerical
findings (cf. Figs \ref{fig1} and \ref{fig2}). 
The loss of coherence is highlighted in Fig. \ref{fig2}, where we
plot the temporal evolution of the modulus squared of an
effective complex order parameter 
measuring the
overall coherence of the system, defined as: 
\begin{equation}
\label{order-parameter}
\Psi=\sum_j \psi_j \psi_{j+1}^{\ast}.
\end{equation}
When the collective oscillations are coherent, the value of the order 
parameter is 
$|\Psi(t)|^2=1$ (see the cases $\xi(t=0)=40,80$ in Fig. \ref{fig2}).
On the other hand, a complete dephasing is characterized by
$|\Psi(t)|^2=0$, and occurs for $\xi(t=0) \ge \xi_{cr}$,
or, equivalently, when $<k> \ge \pi/2$ (cf. Fig. \ref{fig1}). 
It is worth noting that such randomization takes
place between the phases of condensates localized in different wells, 
even though each one remains internally coherent. 
This is shown in Fig. \ref{fig3}, where we plot the density
for different times below (a,b) and above (c,d)
the critical displacement. The numerical solutions of the DNLSE and
the GPE are in good agreement. 
The motion of the center of mass in the supercritical case
is reported in Fig. \ref{fig4}, where the numerical solutions of the DNLS and 
the full one-dimensional 
GPE (\ref{GPE}) are compared. In both cases the system stops 
at $<x> \simeq 35$ $\mu$m (with a slight difference between the
DNLS and the GPE predictions), while the center of the 
harmonic trap is located at $x = 0$ $\mu$m. 
From Eqs. (\ref{phase-current}) we can calculate 
the critical current, i.e.,  the maximum allowed velocity 
in the coherent transmission of matter waves: by setting  
$\Delta \phi=\pi/2$ we readily see that the critical velocity of the center of 
mass, $\dot{\xi}_{max}$ is equal to the critical current per
particle
$I_c=2K/\hbar$. In dimensional units:
\begin{equation}
\label{cr-vel}
v_c=\frac{K \lambda}{\hbar} 
\end{equation}
which gives $v_c=0.98$ $\mu m/ms$,
in agreement with the DNLS numerical result, and close to 
the numerical GPE value $v_c=1.18$ $\mu m/ms$.

From the above findings, we can conclude that the effect of 
the MI is to dephase
the system. In the effective 1D geometry we have considered, 
such dephasing is strong enough to stop the falling condensate. 
In higher dimensions, 
the dephasing can be partial, and will only 
damp the BEC motion. 
Yet, its onset will still be given by Eq. (\ref{cr-vel}). 
The CSIT regards a classical field (the solution of the GPE)
and it is qualitatively different from the quantum Mott insulator-superfluid
transition in mesoscopic Josephson junction chains, which is  
driven by the competition between zero-point
quantum phase fluctuations and the Josephson coupling energy. 
Yet, it is possible to draw an analogy.
In the former CSIT case, the insulator regime is associated with a vanishing
temporal correlation among the phases of each site, each phase still being
meaningful in the GPE sense.
The quantum transition is also driven by 
a loss of phase correlations induced by the localization
of atoms in each site, which, however, arises from the   
non-commuting nature of the number-phase observables. Clearly, such quantum
fluctuations cannot be captured within the GPE framework. 
Also, the latter transition is reversible
(i.e., long-range phase coherence is restored upon adiabatically increase of
the tunnel coupling), while the former is not. It is worth noting
that very recent experimental works \cite{greiner02}
have illustrated the existence
(and reversibility) of the quantum phase transition, rendering
the experimental verification of the classical dynamic transition
suggested herein, a natural next step for experimental studies.
In conclusion, we notice that the MI can be studied in term of
one (or several)
bifurcation points in an effective stationary Hamiltonian.
Such bifurcation points separate regions with different symmetries,
and it is common in the literature to study such dynamical
transitions in terms of
an order parameter \cite{haken}, borrowing the language and concepts
of statistical phase transitions.
This mapping, in the specific case of MI, deserves further studies.

The modulational instability (and the consequent superfluid-insulator
transition) studied here can also be observed with different experimental
setups (e.g., with the 
condensate at the center of the harmonic trap and with 
the laser moving across). In fact, similar MI and pinning results have
been obtained in the case in which the harmonic trap is
displaced at a constant speed exceeding a critical value
\cite{smerzi02}. 
These results illustrate
the generality and importance of the effects of the MI mechanism 
in the motion of Bose-Einstein condensates and underscore its potential
in inducing localization and dephasing of such coherent structures.


\end{multicols}


\begin{thebibliography}{99}
%

\bibitem{anderson98} B.P. Anderson and M.A. Kasevich,
Science {\bf 282},  1686  (1998).

\bibitem{orzel01} C. Orzel, A. K. Tuchman, M. L. Fenselau, M. Yasuda, M. A.
Kasevich, {\it Science} {\bf 291}, 2386 (2001).

\bibitem{morsch01} O. Morsch, J.H. M\"uller, M. Cristiani, D. Ciampini
and E. Arimondo, Phys. Rev. Lett. {\bf 87}, 140402 (2001)

\bibitem{cataliotti01} F.S. Cataliotti, S. Burger, C. Fort, P. Maddaloni, 
F. Minardi, A. Trombettoni, A. Smerzi and M. Inguscio, 
Science {\bf 293}, 843 (2001). 



\bibitem{chiofalo00} M.L. Chiofalo and M.P. Tosi,
\newblock Phys. Lett. A {\bf 268}, 406 (2000).

\bibitem{wu01} B. Wu and Q. Niu,
\newblock Phys. Rev. A {\bf 64}, 061603(R) (2001).

\bibitem{trombettoni01} A. Trombettoni and A. Smerzi,
 Phys. Rev. Lett. {\bf 86}, 2353, (2001).

\bibitem{konotop01} V.V. Konotop and M. Salerno, cond-mat/0106228.

\bibitem{jaksch98} D. Jaksch et al., Phys. Rev. Lett. {\bf 81}, 3108 (1998) 

\bibitem{pedri01} P. Pedri et al., Phys. Rev. Lett. {\bf 87}, 220401 (2001)

\bibitem{benjamin67} T.B. Benjamin and J.E. Feir, 
J. Fluid. Mech. {\bf 27}, 417 (1967).

\bibitem{ostrovskii69} L.A. Ostrovskii, 
Sov. Phys. JETP {\bf 24}, 797 (1969).

\bibitem{taniuti68} T. Taniuti and H. Washimi, 
Phys. Rev. Lett. {\bf 21}, 209 (1968); A. Hasegawa, 
Phys. Rev. Lett. {\bf 24}, 1165 (1970).

\bibitem{peyrard93} M. Peyrard, T. Dauxois, H. Hoyet and C.R. Willis, 
Physica {\bf 68D}, 104 (1993).

\bibitem{morandotti99}  R. Morandotti, U. Peschel, J.S. Aitchison,
\newblock H.S. Eisenberg and Y. Silberberg,
Phys. Rev. Lett. {\bf 83}, 2726 (1999).

\bibitem{kivshar92}  Yu.S. Kivshar and M. Peyrard, 
Phys. Rev. A {\bf 46}, 3198 (1992).

\bibitem{scott99} A.C. Scott, "Nonlinear Science: Emergence and Dynamics
of Coherent Structures", Oxford Univ. Press, Oxford, 1999

\bibitem{hennig99} D. Hennig and G.P. Tsironis, 
Phys. Rep. {\bf 307}, 333 (1999). 

\bibitem{review01} P.G. Kevrekidis, K.{\O}. Rasmussen
and A.R. Bishop,
\newblock Intn. J. Mod. Phys. B {\bf 15}, 2833 (2001).

\bibitem{peyrard84} M. Peyrard and M. Kruskal, 
Physica {\bf 14D}, 88 (1984).

\bibitem{dauxois93} T. Dauxois and M. Peyrard, 
Phys. Rev. Lett. {\bf 70}, 3935 (1993).


\bibitem{kivshar98} Yu.S. Kivshar, D.E. Pelinovsky, 
T. Cretegny and M. Peyrard, Phys. Rev. Lett. {\bf 80}, 5032 (1998). 

\bibitem{kevrekidis00b} P.G. Kevrekidis and C.K.R.T. Jones, 
Phys. Rev. E {\bf 61}, 3114 (2000).


\bibitem{barone} A. Barone and G. Paterno,
$Physics$ $and$ $Applications$ $of$ $the$ $Josephson$ $Effect$
(Wiley, New York, 1982).

\bibitem{kevrekidis00} P.G. Kevrekidis and M.I. Weinstein, 
Physica {\bf 142D}, 113 (2000).

\bibitem{greiner02} M. Greiner {\it et al.},
Nature {\bf 415}, 39 (2002)

\bibitem{haken} H. Haken, ``Synergetics", Springer-Verlag Berlin (1977);
G. Venkataraman and V. Balakrisnan,
"Sinergetica ed Instabilita' dinamiche", Course IC,
Proceedings of the International School of Physics "Enrico Fermi",
Varenna,   Caglioti, Haken and Lugiato Ed.s, (1988), 175

\bibitem{smerzi02}  A. Smerzi, A. Trombettoni, P.G. Kevrekidis
and A.R. Bishop (unpublished).
\end{thebibliography}
\end{document}